\documentclass{webofc}
\usepackage[varg]{txfonts}   % Web of Conferences font
\woctitle{Wigner 111 - Colourful and Deep }
\begin{document}
\title{Equation and test of possible delay time of Newton force}
%
% subtitle is optional

%\author{Lajos Di\'osi\inst{1}\fnsep\thanks{\email{diosi.lajos@wigner.mta.hu}}}
\author{Lajos Di\'osi\thanks{\email{diosi.lajos@wigner.mta.hu}}}

\institute{Wigner Research Center for Physics, H-1525 Budapest 114., P.O.Box 49, Hungary}

\abstract{
Recently, a simple heuristic modification of the Newton potential with a non-zero delay-time
$\tau_G$ has been proposed. Our modification is largely suppressed for purely gravitational 
interactions, it becomes relevant under non-gravitational accelerations of the sources. 
We illustrate how the choice $\tau_G\sim1$ ms may already influence the $5th$ digit of $G$ 
determined by Cavendish experiments. Re-evaluation of old Cavendish experiments and 
implementing slightly modified new ones may confirm the proposal or, at least, put a stronger 
upper limit on $\tau_G$. 
}
\maketitle

\section{Introduction}
\label{sec-0}
Recently, we have discussed a slight non-relativistic modification of the
Newton law of universal gravitation \cite{Dio13}. Alternative to other suggestions 
that modified the $1/r$ shape of the potential \cite{Fisetal86}, the new proposal relaxes
the synchronous emergence of the $1/r$ potential. The emergence takes a certain 
time of the order of $\tau_G$, the Newton force is following the motion of the source
with a certain laziness characterized by the delay time $\tau_G$. The background
motivation came from quantum foundational speculations \cite{Dio14a,Dio14b} yielding an estimate $\tau_G\sim1$ms.
Cavendish-type experiments have extreme low time-resolution, they cannot exclude 
delay times even much greater than $1$ms. This makes our proposal worth of considerations
in itself, independently of the underlying motivation. Sect.~\ref{sec-1} recapitulates
the minimum heuristic modification of the Newton law in order to describe the laziness of the
potential \cite{Dio13}. Sec.~\ref{sec-2} discusses an unexpected effect valid for static sources in Earth's gravity,
Sec.~\ref{sec-3} outlines how this effect can influence the $5th$ digit of the Newton constant $G$ 
determined in Cavendish experiment.

\section{The modified Newton force}
\label{sec-1}
Consider the Newton potential at location $r$ and time $t$, created by a spherically symmetric source $M$ 
at location $x_t$:  
\begin{equation}
\label{Newton}
\Phi(r,t)=-GM\frac{1}{\vert r-x_t\vert}\;,
\end{equation}
where $r$ must be outside the source. 
The simplest modification that contains a delay time $\tau_G$ would take the following form: 
\begin{equation}
\label{Newtonnaive}
\Phi(r,t)=-GM\int_0^\infty\frac{1}{\vert r-x_{t-\tau}\vert}\mathrm{e}^{-\tau/\tau_G}\frac{d\tau}{\tau_G}\;.
\end{equation}
This naive delay equation is not invariant for the Galilean boost of the reference frame,
\begin{equation}
\label{boost}
x_t\Longrightarrow x_t-vt,
\end{equation}
where $v$ is the boost velocity. The boost invariance is restored if in the naive equation we replace 
$x_{t-\tau}$ by $x_{t-\tau}+\dot x_t\tau$. Yet we should satisfy the Newtonian equivalence principle:
the gravitational acceleration $g$ is equivalent with the acceleration $a=g$ of the reference frame:
\begin{equation}
\label{accel}
x_t\Longrightarrow x_t-at^2/2.
\end{equation}
The equivalence principle becomes adopted if we replace $x_{t-\tau}$ by $x_{t-\tau}-g\tau^2/2$. Accordingly, our
ultimate proposal for the modified Newton law takes the following form: 
\begin{equation}
\label{Newtonnew}
\Phi(r,t)=-GM\int_0^\infty\frac{1}{\vert r-x_{t-\tau}-\dot x_t\tau+g\tau^2/2\vert}\mathrm{e}^{-\tau/\tau_G}\frac{d\tau}{\tau_G}\;,
\end{equation}
valid in any inertial frame in the presence of gravity $g$. We assume the field $g$ in the vicinity of the source's locations 
$x_{t-\tau}$ can be considered constant all over the relevant delay period, i.e., for $\tau$ running from $0$ to a few times $\tau_G$.

It is now straightforward to confirm that our proposal is invariant against the boost and acceleration of the reference frame.
We consider the following transformations in (\ref{Newtonnew}): 
\begin{eqnarray}
\label{boostaccel}
x_t&\Longrightarrow&x_t-vt-at^2/2,\\
r  &\Longrightarrow&r  -vt-at^2/2,\\
g  &\Longrightarrow&g-a.
\end{eqnarray}
After elementary steps, we get the original form (\ref{Newtonnew}).
Hence our proposal is invariant for Galilean boosts and accelerations of the reference frame.

The naive non-invariant form (\ref{Newtonnaive}) is not at all useless. We can formulate our proposal
equivalently for it. Observe that the invariant form (\ref{Newtonnew}) reduces to the naive form (\ref{Newtonnaive})
in the instantaneous co-moving free-falling reference frame defined by $v=\dot x_t$ and $a=g$ through (\ref{boostaccel}) since in that frame
$\dot{x}_t=0$ and $g=0$. Accordingly, we can state our proposal like this. The standard Newton law (\ref{Newton}) is 
replaced by the naive delay equation (\ref{Newtonnaive}) which must be evaluated in the instantaneous co-moving free-falling
reference frame. This equivalent interpretation illuminates how we ensure invariance of the mechanism of the delay. 

We can observe a precious bonus of our proposal. As we see, objects free-falling in slowly varying gravity themselves create standard
instantaneous Newton forces (\ref{Newton}). Hence our modification (\ref{Newtonnew}) is irrelevant for purely gravitational many-body 
systems as long as the bodies are distant and/or in slow motion so that they experience largely constant gravitational forces over 
periods of the delay $\tau_G$. 

\section{Effect in Earth's Gravity}
\label{sec-2}
\begin{figure}
\centering
    \includegraphics[width=0.75\textwidth, angle=0]{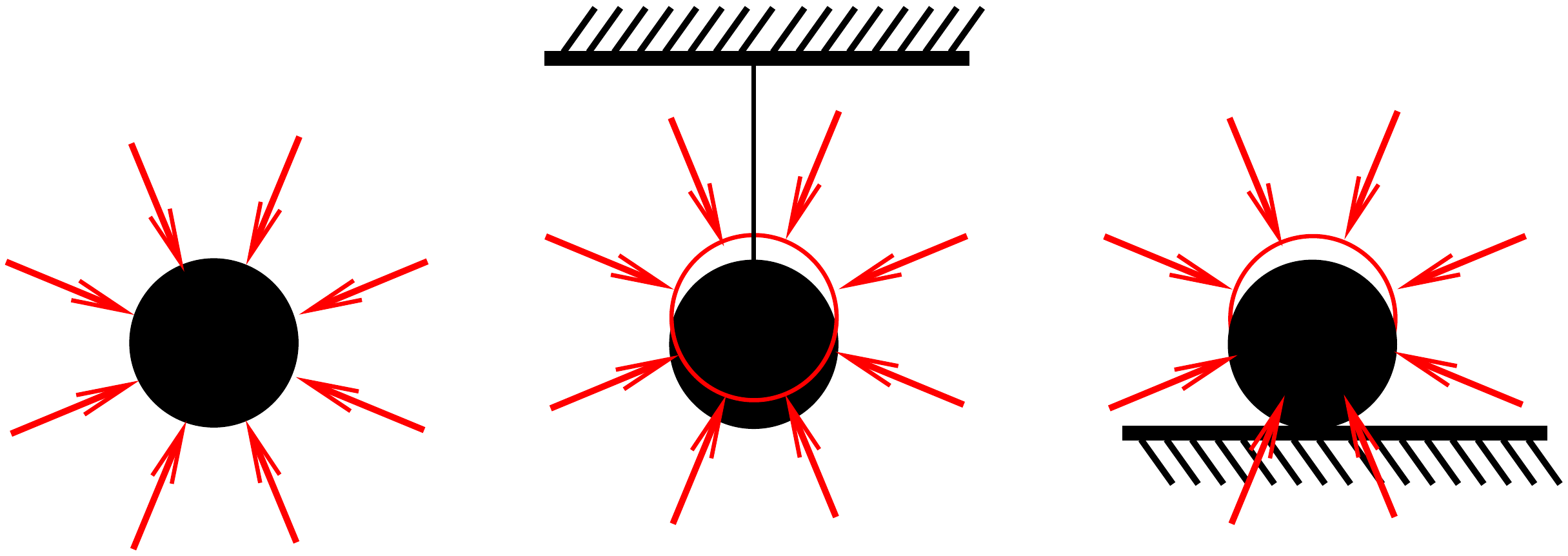}
\caption{(from \cite{Dio14a}) Free-falling source (left) creates standard Newton force.
Static source (middle, right) creates Newton force as if the source were higher than its static geometric location, by $\delta_G=\vert g\vert\tau_G^2~\sim10^{-3}$cm.}
\label{fig-1}
\end{figure}

A surprising consequence of the modified Newton law will occur to all masses resting on Earth \cite{Dio14a}.
Let us first recall that for free-falling objects the proposed law (\ref{Newtonnew}) reduces to Newton's (\ref{Newton}),
the delay mechanism cancels completely (Sec.~\ref{sec-1}).
On the contrary, if the mass is at rest, i.e., it is under the influence of a non-gravitational force $-gM$ where $g$ is Earth's
gravity, then in the instantaneous co-moving free-falling reference system the source is being accelerated upward and we get an 
effect of the delay. Let us calculate it directly from the covariant law (\ref{Newtonnew}). 
Suppose the mass is located at a single point $x_t\equiv x$, hence
\begin{equation}
\label{Newtonnewappl}
\Phi(r,t)=-GM\int_0^\infty\frac{1}{\vert r-x+g\tau^2/2\vert}\mathrm{e}^{-\tau/\tau_G}\frac{d\tau}{\tau_G}\;.
\end{equation}
We see immediately that the effective location of the mass is higher than the geometric location $x$. 
In the lowest order in $g$, we can replace the shift by a constant: 
\begin{equation}
\label{Newtonnewapplshift}
\Phi(r,t)=-GM\frac{1}{\vert r-x+g\tau_G^2\vert}\;.
\end{equation}
With our choice $\tau_G\sim1$ms (Sec.~\ref{sec-0}), the shift itself is
\begin{equation}
\label{deltax}
\delta_G=\vert g\vert\tau_G^2~\sim10^{-3}\mathrm{cm}, 
\end{equation}
and it points upward, cf. figure~\ref{fig-1}.

\section{Effect on Cavendish experiment}
\label{sec-3}
We illustrate the significance of the proposed delay of the Newton force even in simple
Cavendish experiments. Suppose the source and probe masses are spherical symmetric.
Let the horizontal distance be $L=10$cm between the centers of source and probe, respectively.
Assume the geometric locations of the centers are in the same horizontal plane. If, according to
our proposal, the effective positions of the sources are $\delta_G\sim10^{-3}$cm higher, 
the horizontal component of the Newton force on the probe becomes about $1-\frac{2}{3}\delta_G^2/L^2$ times smaller. 
The correction is cca. -0.01 ppm, much less than the related experimental uncertainty of $G$, 
cf. \cite{Quietal13} and references \cite{GunMer00,G} quoted from therein. 

Suppose, however, that the geometric location of the sources are already higher than the locations
of the probes, let the vertical distance of the source w.r.t. the probe be $\delta_0\ll L$. 
This will suitably enhance the effect of $\delta_G$. The horizontal Newton force on the probe becomes about
$1-3\delta_0\delta_G/L^2$ times smaller or larger, depending on the sign of $\delta_0$. 
If the geometric vertical misalignment is as small as $\delta_0=\pm1$mm, 
our proposal predicts cca. $\mp1$ ppm correction to the measured value
of $G$, for $\delta_0=\pm1$ cm the correction is already $\mp10$ ppm, etc. 
\begin{figure}
\centering
    \includegraphics[width=0.75\textwidth, angle=0]{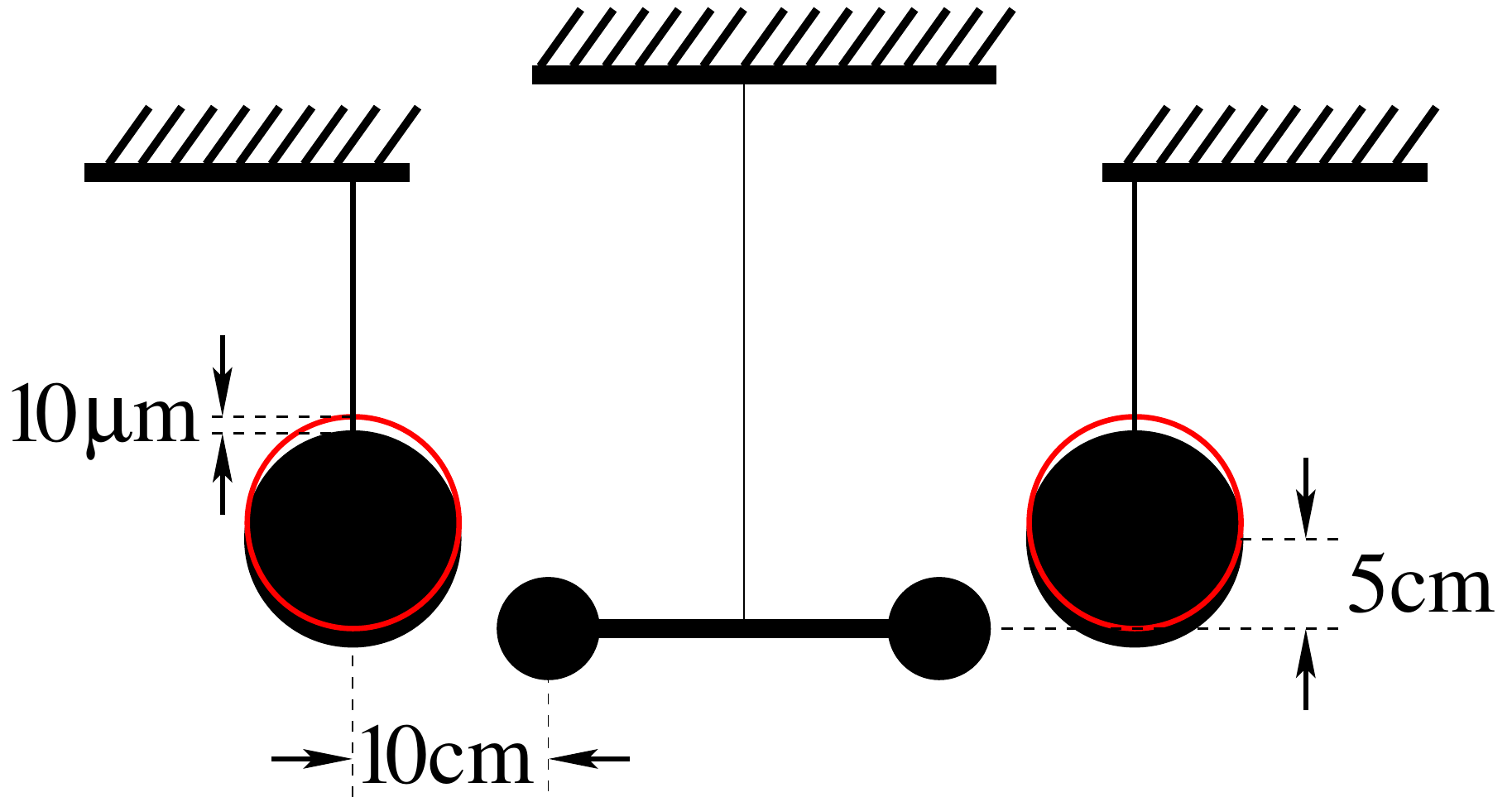}
\caption{Schematic view of a Cavendish experiment where the delay $\tau_G\sim1$ms
would shift the $5th$ digit of the measured $G$ by $-8$.}
\label{fig-2}
\end{figure}
We outline the geometry of a Cavendish experiment where $\delta_0=L/2$, cf. figure~\ref{fig-2}. 
The horizontal Newton force becomes
\begin{equation}
\label{factor}
1-\frac{6}{5}\delta_G/L
\end{equation}
times smaller, yielding -120 ppm correction to the measured $G$. This is already a significant effect
influencing the $5th$ digit of $G$ left so far uncertain notoriously \cite{Quietal13}.  

\section{Remarks}
\label{sec-4}
As we argued in \cite{Dio13}, no experimental evidence exists against our 
scale $\tau_G\sim1$ ms of the delay grounded in quantum foundational speculations \cite{Dio14a,Dio14b}. 
The proposed modification of the Newton law is not a necessary one. Yet, provided it won't 
contain fatal theoretical inconsistency, we should confirm or refute it experimentally.
It is anyway desirable that experiments put an upper limit on the delay time $\tau_G$.
There are various predictions that are detectable in principle \cite{Dio14a}. 
Here we have suggested the one feasible in Cavendish experiments even without time-resolution 
of the delay itself.  

\begin{acknowledgement}
This work was supported by the Hungarian Scientific Research Fund under Grant No. 75129 and
the EU COST Action MP1006 `Fundamental Problems in Quantum Physics'. 
\end{acknowledgement}


\begin{thebibliography}{99}
\bibitem{Dio13} L. Di\'osi, Phys. Lett. A {\bf 377}, 1782 (2013); 
Erratum cited from \cite{Dio14a}: redundant $1/2$ in (4).  
\bibitem{Fisetal86} E. Fischbach, D. Sudarsky, A. Szafer, C. Talmadge, S. H. Aronson, Phys. Rev. Lett. {\bf 56}, 3 (1986)
\bibitem{Dio14a} L. Di\'osi, J. Phys. Conf. Ser. {\bf 504}, 012020 (2014)
\bibitem{Dio14b} L. Di\'osi, Found. Phys. {\bf 44}, 483 (2014)
\bibitem{Quietal13} T. Quinn, H. Parks, C. Speake, R. Davis, Phys. Rev. Lett. {\bf 111}, 101102 (2013)
\bibitem{GunMer00} J. H. Gundlach, S. M. Merkowitz Phys. Rev. Lett. {\bf 85}, 2869 (2000)
\bibitem{G} G. G. Luther, W. R. Towler, Phys. Rev. Lett. {\bf 48}, 121 (1982); 
            O. V. Karagioz, V. P. Izmailov, Meas. Tech. {\bf 39}, 979 (1996); 
            C. H. Bagley, G. G. Luther, Phys. Rev. Lett. {\bf 78}, 3047 (1997); 
            U. Kleinvoss, Ph.D. thesis, University of Wuppertal, 2002;
            T. R. Armstrong, M. P. Fitzgerald, Phys. Rev. Lett. {\bf 91}, 201101 (2003); 
            Z.-K. Hu, J.-Q. Guo, J. Luo, Phys. Rev. D {\bf 71}, 127505 (2005); 
            S. Schlamminger et al., 
                                    %E. Holzschuh, W. K\"undig, F. Nolting, R. Pixley, J. Schurr, U. Straumann, 
                                    Phys. Rev. D {\bf 74}, 082001 (2006); 
            J. Luo et al., %Q. Liu, L.-C. Tu, C.-G. Shao, L.-X. Liu, S.-Q. Yang, Q. Li, Y.-T. Zhang, 
                           Phys. Rev. Lett. {\bf 102}, 240801 (2009);
            H. V. Parks, J. E. Faller, Phys. Rev. Lett. {\bf 105}, 110801 (2010)
\end{thebibliography}
\end{document}